\documentclass[prb,twocolumn,showpacs,preprintnumbers,amsmath,amssymb]{revtex4}
\usepackage{graphicx}
\usepackage{dcolumn}
\usepackage{bm}
\usepackage{natbib}
\usepackage[T1]{fontenc}
\begin{document}

\title{Magnetic phase diagram for a non-extensive system:
Experimental connection with manganites}

\author{M.S. Reis}
\altaffiliation[Present address: ]{Centro Brasileiro de Pesquisas
Físicas, Rua Dr. Xavier Sigaud 150 Urca, 22290-180 Rio de
Janeiro-RJ, Brasil} \email{marior@fis.ua.pt}
\author{V.S. Amaral}
\affiliation{Departamento de Física, Universidade de Aveiro,
3810-193 Aveiro, Portugal}
\author{J.P. Araújo}
\affiliation{IFIMUP, Departamento de Física, Universidade do
Porto, 4150 Porto, Portugal}
\author{I.S. Oliveira}
\affiliation{Centro Brasileiro de Pesquisas Físicas, Rua Dr.
Xavier Sigaud 150 Urca, 22290-180 Rio de Janeiro-RJ, Brazil}

\date{\today}

\begin{abstract}
In the present paper we make a thorough analysis of a classical
spin system, within the framework of Tsallis nonextensive
statistics. From the analysis of the generalized Gibbs free
energy, within the mean-field approximation, a para-ferromagnetic
phase diagram, which exhibits first and second order phase
transitions, is built. The features of the generalized, and
classical, magnetic moment are mainly determined by the values of
$q$, the non-extensive parameter. The model is successfully
applied to the case of La$_{0.60}$Y$_{0.07}$Ca$_{0.33}$MnO$_3$
manganite. The temperature and magnetic field dependence of the
experimental magnetization on this manganite are faithfully
reproduced. The agreement between rather "exotic" magnetic
properties of manganites and the predictions of the
$q$-statistics, comes to support our early claim that these
materials are magnetically nonextensive objects.
\end{abstract}


\maketitle

\section{Introduction}\label{section_Introduction}

In the literature of manganites, various models have appeared as
different attempts to reproduce the electric and magnetic
properties of these systems. Krivoruchko \textit{et al.}
\cite{JMMM_207_1999_168}, Nunez-Regueiro \textit{et al.}
\cite{APL_68_1996_2747} and Dionne \cite{JAP_79_1996_5172} are
interesting examples of multiparameter models, but which failed to
achieve full agreement to experimental data. Ravindranath
\textit{et al.} \cite{PRB_63_2001_184434} compare resistivity
data in La$_{0.6}$Y$_{0.1}$Ca$_{0.3}$MnO%
$_3$ to different two-parameter models, which do not agree to each
other in the low-temperature range. Other interesting attempts can
be found in the work of Rivas \textit{et al.}
\cite{JMMM_221_2000_57}, Hueso \textit{et al.}
\cite{JMMM_189_1998_321}, Heremans \textit{et al.}
\cite{JAP_81_1997_4967} , Pal \textit{et al.}
\cite{JAP_89_2001_4955}, Philip \textit{et al.}
\cite{JPCM_11_1999_8537}, Szewczyk \textit{et al.}
\cite{APL_77_2000_1026}, Viret \textit{et al.}
\cite{PRB_55_1997_8067} and Tkachuk \textit{et al.}
\cite{PRB_57_1998_8509}. None of these obtained plain agreement
between experiment and theory, irrespect their number of adjusting
parameters and approach.

On another hand, Tsallis generalized statistics
\cite{JSP_52_1988_479,{PA_261_1998_534},{livro_tsallis},{BJP_29_1999_1}},
has been successfully applied to an impressive number of areas
\cite{website}. The formalism rests on the definition of
generalized entropy \cite{JSP_52_1988_479}:
\begin{equation} \label{entropia}
S_{q}=k \frac{1-\sum_{i}p_{i}^{q}}{q-1}
\end{equation}
where $q$ is the {\em entropic index}, $p_{i}$ are probabilities
satisfying $\sum_{i}p_{i}=1$ and $k$ is a positive constant. The
above formula converges to the usual Maxwell-Boltzmann definition
of entropy, and to the usual derived thermodynamic functions,  in
the limit $q\rightarrow 1$
\cite{JSP_52_1988_479,{BJP_29_1999_1},{PA_261_1998_534},{livro_tsallis},{website}}.

In what concerns Condensed Matter problems, applications of
Eq.\ref{entropia} include: Ising ferromagnets
\cite{PA_175_1991_285,{PA_203_1994_486},{APPA_91_1997_1035},{JPA_30_1997_3345},{PRB_55_1997_5611}},
molecular field approximation
\cite{TJP_21_1997_132,{PRB_61_2000_11521},{PA_238_1997_285}},
percolation problems \cite{PA_266_1999_42}, Landau diamagnetism
\cite{EPJB_2_1998_101,{ZPB_104_1997_341}}, electron-phonon systems
and tight-binding-like Hamiltonians
\cite{PRE_55_1997_7759,{PRB_60_1999_4629},{BJP_29_1999_169}},
metallic \cite{EPJB_14_2000_43} and superconductor
\cite{PA_296_2001_106} systems, etc. The first evidences that the
magnetic properties of manganites could be described within the
framework of Tsallis statistics were presented by Reis \emph{et
al.} \cite{EL_58_2002_42}, followed by an analysis
\cite{PRB_66_2002_134417} of the unusual paramagnetic
susceptibility of La$_{0.67}$Ca$_{0.33}$MnO$_3$, measured by
Amaral \emph{et al.}\cite{JMMM_226_2001_837}.

Maximization of Eq.\ref{entropia} subjected to the constraint of
the normalized $q$-expectation value of the Hamiltonian
$\mathcal{\hat{H}}$ \cite{{PA_261_1998_534},{livro_tsallis}}:
\begin{equation} \label{energia_interna}
  U_q=\frac{Tr\{\mathcal{\hat{H}}\hat{\rho}^q\}}{Tr\{\hat{\rho}^q\}}
\end{equation}
and the usual normalization of the density matrix
$Tr\{\hat{\rho}\}=1$, yields the following expression for the
density matrix $\hat{\rho}$:
\begin{equation} \label{rho}
\hat{\rho}
=\frac{1}{Z_{q}}[1-(1-q)\tilde{\beta}({\mathcal{\hat{H}}}-U_{q})]^{1/(1-q)}
\end{equation}
where
\begin{equation}\label{part_function}
Z_{q}=Tr[1-(1-q)\tilde{\beta}({\mathcal{\hat{H}}}-U_{q})]^{1/(1-q)}
\end{equation}
is the partition function and
$\tilde{\beta}=\beta/Tr\{\hat{\rho}^{q}\}$. Here, $\beta$ is the
Lagrange parameter associated to the internal energy.

The magnetization of a specimen is, accordingly, given by
\cite{EL_58_2002_42,{PRB_66_2002_134417},{livro_tsallis},{PA_261_1998_534},{BJP_29_1999_1}}:
\begin{equation}\label{mag_qexpect_value}
M_{q}=\frac{Tr\{\hat{\mu} \hat{\rho} ^{q}\}}{Tr\{\hat{\rho}
^{q}\}}
\end{equation}
where $\hat{\mu}$ is the magnetic moment operator.

However, Eq.\ref{rho} can be written in a more convenient form
\cite{PA_261_1998_534,{EL_58_2002_42},{livro_tsallis},{PRB_66_2002_134417}}
in terms of $\beta^{\ast}$, defined as
$\beta^{\ast}=\tilde{\beta}/[1+(1-q)\tilde{\beta}U_{q}]$. In
particular, to analyze the physical system described here, the
quantity $1/(k \beta ^{\ast })$ will represent the physical
temperature scale, as in references
\cite{EL_58_2002_42,{PRB_66_2002_134417}}. Discussion about the
concept of temperature and Lagrange parameters in Tsallis
statistics can be found in the literature
\cite{PA_261_1998_534,{livro_tsallis},{livro_rajagopal},{PLA_281_2001_126},{PA_300_2001_417},{EL_58_2002_42},{PRB_66_2002_134417}}.
In the present work, the $q$ parameter is restricted to the
interval $0\leq q \leq1$, preserving the entropy concavity
\cite{PRL_83_1999_1711, {PRB_66_2002_134417}}.

In this paper we pursue the idea, based on novel experimental and
theoretical results, that manganites are magnetically
non-extensive objects. This property appears in systems where
long-range interactions and/or fractality exist, and such features
have been invoked in recent models of manganites, as well as in
the interpretation of experimental results. They appear, for
instance, in the work of Dagotto and co-workers
\cite{PR_344_2001_1}, who emphasize the role of the competition
between different phases to the physical properties of these
materials. Various authors have considered the formation of
micro-clusters of competing phases, with \emph{fractal} shapes,
randomly distributed in the material
\cite{PRL_86_2001_135,{PRB_59_1999_7033}}, and the role of
\emph{long-range interactions} to phase segregation
\cite{science_283_1999_2034,{PRB_64_2001_235127}}. Important
experimental results in this direction have also been reported by
Marithew \textit{et al.} \cite{PRL_84_2000_3442}, and Fiebig
\textit{et al.} \cite{science_280_1998_1925}. Particularly
insightful is the work of Satou and Yamanada
\cite{PRB_63_2001_212403}, who derived a \emph{Cantor} spectra for
the double-exchange hamiltonian, basis of theoretical models of
manganites.

A major difficulty with Tsallis formulation concerns the physical
meaning of the entropic parameter $q$. In this direction, Beck and
Cohen \cite{cond-mat/0205097} have recently shown that the value
of $q$ gives a direct measure of the internal distribution of
temperatures in an \emph{inhomogeneous} system. Although their
results are not directly applied to magnetic systems, it has been
known for some time that manganites are magnetically inhomogeneous
systems (see, for instance, Ref.\cite{PR_344_2001_1,{NMR_prca}}
and references therein), and this fact has been explored very
recently by Salamon \emph{et al.} \cite{PRL_88_2002_197203}, who
applied the idea of \emph{distribution} of the inverse
susceptibility (which turns out to be equivalent to a distribution
of temperatures), to the analysis of the magnetic susceptibility
and the effective paramagnetic moment of
La$_{0.7}$Ca$_{0.3}$MnO$_3$.

In what follows, we present a magnetic model for classical spins
(cluster), using for that, the Tsallis generalized statistic. In
the model, we consider that non-extensivity exists in the
intra-cluster interaction, whereas the inter-cluster interaction
remains extensive. This is important to maintain the total
magnetization proportional to the number of clusters. Following,
the Gibbs free energy is analyzed, within the mean-field
approximation, and a series of interesting magnetic features
appears, as a consequence of the intra-cluster non-extensivity.
Finally, a connection between the model proposed here and the
experimental data obtained from magnetic measurements performed on
manganites support our thesis that these objects are magnetically
non-extensive.

\section{Classical Model}\label{section_Classical_Model}

Consider a classical spin $\vec{\mu}$ submitted to a homogeneous
magnetic field $\vec{H}$. The hamiltonian $\mathcal{H}$ is given
by
\begin{equation}\label{hamiltoniana}
  \mathcal{H}=-\mu H \cos\theta
\end{equation}
where $\theta$ is the angle between $\vec{\mu}$ and $\vec{H}$.
Following the usual Tsallis formalism
\cite{EL_58_2002_42,{PRB_66_2002_134417},{livro_tsallis},{PA_261_1998_534},{BJP_29_1999_1}},
the magnetization $\mathcal{M}_q$ can be determined from
Eq.\ref{mag_qexpect_value}, yielding:
\begin{equation}\label{langevin_general}
\frac{\mathcal{M}_q}{\mu }=\mathcal{L}_{q}\left( x\right)
=\frac{1}{\left( 2-q\right) } \left\{
\begin{array}{cc}
1-\frac{1}{x}, & x>\frac{1}{1-q} \\
\coth _{q}(x)-\frac{1}{x}, & x<\frac{1}{1-q}
\end{array}
\right.
\end{equation}
where $x=\mu H/kT$, and $coth_q$ is the generalized q-hyperbolic
co-tangent \cite{JPAMG_31_1998_5281}. The above two branches
function  results from the Tsallis cut-off
\cite{livro_tsallis,{PLA_256_1999_221},{BJP_29_1999_50},{BJP_29_1999_1}}.
It is interesting to note the similarity between the above result
and the traditional Langevin function. In what follows,
Eq.\ref{langevin_general} will be called \emph{Generalized
Langevin Function}, and this result can also be derived from the
\emph{Generalized Brillouin Function}, introduced in
Ref.\cite{PRB_66_2002_134417}, taking the limit of large spin
values ($S \rightarrow \infty$). The result derived above is valid
only for 0$\leq q\leq $1.

The generalized magnetic susceptibility $\chi_q$ shows the usual
dependence on the inverse of the absolute temperature
\begin{equation}\label{sus_para}
  \chi_q=\lim_{H\to\\0} \left[\frac{\partial\mathcal{M}_q}{\partial H}
  \right]=\frac{q \mu^2}{3kT}=q \chi_{\textnormal{\tiny{1}}}
\end{equation}
and is proportional to the usual paramagnetic Langevin
susceptibility $\chi_{\textnormal{\tiny{1}}}$. A similar result
was deduced for the generalized Brillouin function
\cite{PRB_66_2002_134417}.

\section{Mean-Field Approximation}\label{section_Mean-Field_Approximation}

\subsection{Gibbs Free Energy}\label{subsection_Gibbs_free_energy}

The Gibbs free energy is, within the mean field approximation,
\begin{equation}\label{gibbs}
  G=\frac{kT}{\mu}\int^{\mathcal{M}_q}_0 \mathcal{M}^{-1}_q(\mathcal{M}'_q)
   d\mathcal{M}'_q -H\mathcal{M}_q - \frac{\lambda}{2}\mathcal{M}_q^2
\end{equation}
where the first term is the entropy, with $\mathcal{M}_q^{-1}$ the
inverse function of the generalized Langevin function
(Eq.\ref{langevin_general}); the second is the Zeeman term; and
the third, the exchange energy, with $\lambda$ as the mean field
parameter. The equilibrium magnetization can be found from the
minimization of the Gibbs free energy and, depending on the $q$
value, first or second order transition features emerge. At zero
magnetic field and for q$>$0.5 the free energy presents, at any
temperature, only one minimum, for positive values of
magnetization. Correspondingly, a second order para-ferromagnetic
phase transition occurs at $T_c^{(q)}=qT_c^{(1)}$, as sketched in
figure 1(a). Here, $T_c^{(1)}=\mu^2 \lambda/3k$ is the Curie
temperature for the standard Langevin model. From now on, we
introduce the dimensionless temperature and field parameters
$t=T/T_c^{(1)}$ and $h=\mu H/kT_c^{(1)}$.

For q$\leq$0.5 the nature of the phase transition is more complex,
presenting a typical behavior of first order phase transition
\cite{livro_transicao_de_fase}. As it is illustrated in figure
1(b), for sufficiently high temperatures ($t>t_{SH}$), only one
minimum at $\mathcal{M}_q=0$ is observed. Decreasing temperature,
at $t=t_{SH}$, a second minimum appears with finite magnetization
and further lowering temperature to $t=t_c$, this minimum becomes
degenerate, corresponding to $\mathcal{M}_q=0$. For $t<t_{c}$ the
free energy global minimum occurs for finite magnetization,
determining its equilibrium value. However, this equilibrium state
is not necessarily the one observed in finite times, since there
is an energy barrier between the two minima, which prevent the
whole system to reach the global minimum of energy. The minimum
temperature that can sustain zero magnetization is the one
corresponding to $t=t_{SC}$, where the energy barrier goes to zero
and only one minimum exists at finite magnetization. In a similar
way, $t_{SH}$ corresponds to the maximum temperature that can
sustain a finite magnetization. These phenomena are the well-known
`superheating' and `supercooling' cycles and are responsible for
the thermal hysteresis normally observed in first order phase
transitions \cite{livro_transicao_de_fase}. The $t_{SH}$ and
$t_{SC}$ temperatures can be analytically derived from the
conditions described above, and are valid only for $q\leq$0.5,
yielding:
\begin{equation}\label{tsh}
t_{SH}=\frac{3}{4}\frac{1}{(2-q)}
\end{equation}
\begin{equation}\label{tsc}
  t_{SC}=q
\end{equation}

On the contrary, a closed expression for $t_c$  cannot be derived,
since it involves transcendental equations. The temperature
dependence of the reduced equilibrium magnetization $\mathcal{M}_q
(2-q)/\mu$ is sensitive to the features described above, as it can
be seen in figure 2 and inset therein.

\subsection{Magnetic Susceptibility}\label{subsection_Magnetic_susceptibility}

The generalized magnetic susceptibiliy $\chi_q$, for any $q$
value, can be derived from Eq.\ref{gibbs},
\begin{equation}\label{sus}
  \chi_q=\frac{C^{(q)}}{t-\theta_p^{(q)}}
\end{equation}
where, in analogy with the usual Curie-Weiss law, we define
\begin{equation}\label{const_curie}
  C^{(q)}= \frac{\mu}{k T_C^{(1)}}\left [\left. \frac{\partial \mathcal{M}_q^{-1}}
 {\partial \mathcal{M}_q}\right|_{\mathcal{M}_q(H=0,T)}\right ]^{-1}
\end{equation}
as the \emph{Generalized Curie Function}. In the paramagnetic
phase, $t\geq t_c$, $M(H=0,T)=0$ and the functions $C^{(q)}$ and
$\theta_p^{(q)}=\lambda C^{(q)}$ are constants:
\begin{equation}\label{curie_generalizada_tmaiortc}
  C^{(q)}=\frac{q}{\lambda}
\end{equation}
\begin{equation}\label{tetap_generalizada_tmaiortc}
  \theta_p^{(q)}=q
\end{equation}

The temperature dependence of $\chi_q$, and its inverse, are quite
distinct, depending whether $q<0.5$ or $q>0.5$, as displayed in
figure 3(a) and (b), respectively. For $q>0.5$ the susceptibility
diverges at $t_c$, the same temperature where its inverse
intercepts the temperature axis at $\theta_p^{(q)}$. For $q<0.5$
the susceptibility is always finite and peaks discontinuously at
$t_c$. Correspondingly, its inverse shows a discontinuity with
finite values and the Curie-Weiss linear behavior extrapolates to
$\theta_p^{(q)}$, which is lower than $t_c$.

\subsection{Influence of Magnetic Field on the Phase Transition}\label{subsection_Influence_of_magnetic_field}

From the analysis of the Gibbs free energy, we conclude that for
$q<0.5$ and $h=0$, the equilibrium magnetization has a first order
phase transition. Sufficiently high magnetic field $h\geq h_{0q}$,
is able to remove the energy barrier between the two minima in the
Gibbs free energy and, consequently, the discontinuity in the
equilibrium magnetization curve. This effect is illustrated in
figure 4. The expression for $h_{0q}$ can be derived from the
condition described above, yielding:
\begin{equation}\label{b0}
  h_{0q}=\frac{3(1-2q)}{2-q}
\end{equation}

The quantity $h/\mathcal{M}_q$ is particularly important from the
experimental point of view, and is sketched in figure 5, for q<0.5
and $h=3h_{0q}$.

Another interesting point of this model, for $q<0.5$, is the
existence of a \emph{ferromagnetic phase transition induced by the
magnetic field}, in the paramagnetic phase. In fact, sufficiently
close to $t_c$, the two minima have similar free energy values
and, consequently, the magnetic field energy plays a decisive
role, being able to switch between the low and high magnetization
values, inducing the first order phase transition. However, for
sufficiently high temperatures $t\geq t_{0q}$, only one minimum
exists for any value of magnetic field, and the phase transition
becomes of second order character. This effect is illustrated in
figure 6. The temperature $t_{0q}$, which determines the
continuous or discontinuous nature of the magnetization versus
field curve is given by:
\begin{equation}\label{t0}
  t_{0q}=\frac{3(1-q)^2}{2-q}
\end{equation}

Above $t_{0q}$, the magnetization as a function of magnetic field
curves are continuous, presenting, however, a characteristic
change of slope at such magnetic field $h_c$, which varies
linearly with temperature, and is given by
\begin{equation}\label{Bc}
  h_c=\alpha (t-t')
\end{equation}

where
\begin{equation}\label{t'}
  t'=\frac{3q(1-q)}{(2-q)}
\end{equation}

and
\begin{equation}\label{alfa}
  \alpha=\frac{1}{1-q}
\end{equation}

\subsection{Magnetic Phase Diagrams}\label{subsection_magnetic_phase_diagrams}

At this point it is convenient to summarize the main properties
found so far: the behavior of the generalized magnetization and
the generalized magnetic susceptibility with temperature and field
depend greatly on the values assumed for the entropic parameter
$q$. It can be the usual second order para-ferromagnetic phase
transition but it can become first order, exhibiting the
properties normally associated to this type of transitions.

Figure 7(a) presents the $t-q$ phase diagram for several $h$
values. The solid lines divide the plane in ferro (below) and
paramagnetic (above) regions. For $h=0$ and $q>0.5$ the
para-ferromagnetic phase transition is always second order,
whereas for $q<0.5$ the transition becomes first order. The dotted
lines limit the stability regions of the fully ordered and
disordered phases. In this way, the shaded area between such
dotted lines represents the ferro-paramagnetic \emph{phase
coexistence}, which can exist for temperatures in the interval
$t_{SC}(q)\leq t \leq t_{SH}(q)$.

Strictly, in the presence of the magnetic field $h$ the
ferro-paramagnetic phase transition does not exist for any
temperature, since $\mathcal{M}_q$ is always different from zero.
In any case, for $q<0.5$, if the magnetic field is not too high,
the Gibbs free energy presents two minima close to $t_c$ and we
can always distinguish two phases, one with small magnetization
$\mathcal{M}_q^{small}$ and other with large magnetization
$\mathcal{M}_q^{large}$. If the magnetic field is sufficiently
high $h \geq h_{0q}$, the energy barrier disappears and the
transition becomes continuous. However, in this case, the
transition occurs with a characteristic slope change in
magnetization versus temperature curves, which does not occur for
$q>0.5$. For  $q<0.5$ and up to a critical applied magnetic field,
the phase transition occurs discontinuously, with the shaded areas
corresponding to the values of temperature and $q$ where phase
coexistence occurs. The magnitude of the discontinuity $\Delta
\mathcal{M}_q$ decreases with increasing field, disappearing for
$h\geq h_{0q}$ (Eq.\ref{b0}), above which the transition is always
second order (see figure 4). The dashed line cuts the transition
lines at the point ($q_{0h},t_{0h}$), which divides each line into
a first and second order region. As the magnetic field increases,
this points travels in the dashed curve according to the
parametric equations:
\begin{equation}\label{q0h}
  q_{0h}=\frac{3-2h}{6-h}
\end{equation}
\begin{equation}\label{t0h}
  t_{0h}=\frac{1}{3}\frac{(3+h)^2}{6-h}
\end{equation}

In an analogous way, figure 7(b) presents the projection of the
phase diagram in the $h-q$ plane for several temperatures $t>t_c$.
Each line divides the plane in the ferromagnetic region (above the
curve) and in the paramagnetic region (below the curve). For
$q<0.5$ and sufficiently close to $t_c$, a field induced phase
transition occurs discontinuously, with the dashed areas
corresponding to the regions of phase coexistence. The magnitude
of the discontinuity $\Delta \mathcal{M}_q$ decreases with
increasing temperature, disappearing for $t\geq t_{0q}$
(Eq.\ref{t0}), above which the transition is always second order
(see figure 6). The dashed line cuts the transition curves at the
point $(q_{0t},h_{0t})$, which similarly as in the previous
diagram, divides the curves in regions of first and second order
phase transition. As the temperature increases, the point moves up
along the curve according to the parametric equations given below,
with corresponding increase of the second order region:
\begin{equation}\label{q0h}
  q_{0t}=\frac{6-t-\sqrt{t^2+12t}}{6}
\end{equation}
\begin{equation}\label{t0h}
  h_{0t}=\frac{6(t-3+\sqrt{t^2+12t})}{t+6+\sqrt{t^2+12t}}
\end{equation}

The projection of the phase diagram in the $h-t$ plane is
presented in figure 7(c), only for $q$=0.1, for sake of clearness.
Again, the dotted lines limit the stability regions of the para
and ferromagnetic phases and the shaded area represents the values
of field and temperature where the two phases can coexist. Above
certain values of field and temperature $h_{0q}$ and $t_{0q}$, the
transition becomes second order and the characteristic field $h_c$
varies linearly with temperature, as given by Eq.\ref{Bc}. The
open circle $(t_{0q},h_{0q})$ divides the curve into the first and
second order phase transition regions, and travels along the
dashed curve according to Eqs.\ref{b0} and \ref{t0}.

\subsection{Generalized Landau Coefficients}\label{section_generalized_landau_coef}

In this section we derive the generalized coefficients of Landau
theory of phase transitions \cite{livro_transicao_de_fase}. Let us
assume a reduced Gibbs free energy $\mathcal{G}=G/kT_C^{(1)}$, and
magnetization $m=\mathcal{M}_q/\mu$. Thus, we are able to expand,
for small values of magnetization, the previously presented Gibbs
free energy (Eq.\ref{gibbs}), yielding:
\begin{equation}\label{gibbis_rescaled}
    \mathcal{G}=
    \frac{\mathcal{A}_q}{2}m^2+\frac{\mathcal{B}_q}{4}m^4+\frac{\mathcal{C}_q}{6}m^6-hm
\end{equation}
where,
\begin{equation}\label{Aq}
    \mathcal{A}_q=\frac{3}{q}(t-q)
\end{equation}
\begin{equation}\label{Bq}
    \mathcal{B}_q=\frac{9(8q-3-4q^2)t}{5q^3}
\end{equation}
\begin{equation}\label{Cq}
    \mathcal{C}_q=\frac{27(54-318q+623q^2-464q^3+116q^4)t}{175q^5}
\end{equation}

Within the Landau theory, negative values of the coefficient
$\mathcal{B}$ mean first order transitions. In this direction, an
obvious correlation between the model here proposed and the usual
Landau theory is obtained, since for $q<$1/2, the
\emph{generalized Landau coefficient} $\mathcal{B}_q$ also assume
negative values, as sketched in figure 8. Note that for $q<$1/2
our model also predicts first order transition. Additionally,
still within the Landau theory, the parameter $\mathcal{A}$
usually takes the form $\mathcal{A}=a(T-T_0)$ (Curie Law), and
this is exactly the relation found for the \emph{generalized
Landau coefficient} $\mathcal{A}_q$ (Eq.\ref{Aq} and sketched on
the inset of figure 8), that represents the inverse of the
generalized susceptibility (Eq.\ref{sus}).

It is well know
\cite{PR_126_1962_104,{PL_12_1964_16},{PB_320_2002_23}} that,
using the classical formulation of the Landau theory, or similar,
a negative slope of the isotherm plots $h/m$ vs. $m^2$ (Arrot
Plot) would indicate a first order phase transition. Thus,
deriving the minimum of the Gibbs free energy
($d\mathcal{G}/dm=0$), we can express the $h/m$ quantity as:
\begin{equation}\label{arrot}
    \frac{h}{m}=\mathcal{A}_q+\mathcal{B}_q m^2+\mathcal{C}_q
    (m^2)^2
\end{equation}

As expected, for $q<$1/2 the \emph{Generalized Arrot Plot} has a
negative slope, indicating first order transition, whereas for
$q>$1/2, these plots are straight lines, characteristic of a
ferromagnetic second order phase transition. These features are
presented in figure 9(a) and (b), for $q>$1/2 and $q<$1/2,
respectively.

\section{Connections to Experimental Results}\label{section_Connections_to_exp.}

\subsection{A Brief Survey} \label{subsection_overview}

The experimental field and temperature dependencies of some
manganites present interesting aspects. Mira \emph{et al.}
\cite{PRB_60_1999_2998,{PB_320_2002_23}} analyzed the character of
the phase transition in
La$_{2/3}$(Ca$_{1-y}$Sr$_y$)$_{1/3}$MnO$_3$ and concluded that for
y=0 the magnetic transition is of first order, whereas it is
second order for y=1. Other works, including those using nuclear
magnetic resonance (NMR), support this result
\cite{PRB_60_1999_6655,{PRB_63_2000_024421}}. Amaral \emph{et al.}
\cite{JAP_sem_numero,{JMMM_226_2001_837},{JMMM_242_2002_655}}
emphasized that La$_{0.67}$Ca$_{0.33}$MnO$_3$, La$_{0.8}$MnO$_3$
and La$_{0.60}$Y$_{0.07}$Ca$_{0.33}$MnO$_3$ exhibit first order
transition character, with additional hysteresis for fields below
than a critical field H$<$H$_c^*$ and temperature ranges between
T$_C$ and a critical temperature T$_C^*$. For the last manganite
cited above, for instance, M(H) presents an upward inflection
point from T$_C$=150 K up to ~220 K, with a characteristic field
H$_c$(T), for the inflexion point, presenting an almost liner
temperature dependence. In addition, a large thermal hysteresis is
clearly delimited for temperatures ranging from T$_C$ up to the
branching point T$_C^*\sim$170 K. Analogous behavior are found in
La$_{0.67}$Ca$_{0.33}$MnO$_3$ and La$_{0.8}$MnO$_3$
\cite{JAP_sem_numero},
Sm$_{0.65}$Sr$_{0.35}$MnO$_3$\cite{PRB_60_1999_12847} and
Pr$_{0.5}$Ca$_{0.5}$Mn$_{0.95}$Cr$_{0.05}$O$_3$
\cite{SSC_114_2000_429}, among others.

Another interesting feature on the magnetization behavior of some
manganites concerns the H/M vs T measurement, that presents a
strong downturn for temperatures nearly above T$_C$. It makes a
deviation from the simple Curie-Weiss law, and, at T$_C$, the
magnetic state is quickly switched to a ferromagnetic one. Such
feature is frequently found in the literature of manganites:
La$_{0.60}$Y$_{0.07}$Ca$_{0.33}$MnO$_3$ and La$_{0.8}$MnO$_3$
\cite{JMMM_242_2002_655}, Sm$_{0.55}$Sr$_{0.45}$MnO$_3$
\cite{JMMM_200_1999_1,{PRB_65_2002_100403}},
La$_{0.825}$Sr$_{0.175}$Mn$_{0.86}$Cu$_{0.14}$O$_3$
\cite{PRB_62_2000_1193}, Ca$_{1-x}$Pr$_{x}$MnO$_3$ with $x\leq$0.1
\cite{PRB_62_2000_9532}, among others.

However, even with the enormous quantity of experimental
measurements on manganites (see, for example, the impressive
review by Dagotto and co-workers \cite{PR_344_2001_1}), the nature
of the phase transition in ferromagnetic manganites is still a
controversial issue. In this direction, Amaral \emph{et al.}
\cite{JAP_sem_numero,{JMMM_242_2002_655}} claim for new models to
explain the first-second order character of the transition in
manganites, even with the theoretical works developed by Jaime
\emph{et al.} \cite{PRB_60_1999_1028}, Alonso \emph{et al.}
\cite{PRB_63_2001_054411} and Novák \emph{et al.}
\cite{PRB_60_1999_6655}.

To inquire about the order of the phase transition and describe
theoretically the behavior of the relevant magnetic quantities,
Amaral and co-workers \cite{JAP_sem_numero,{JMMM_242_2002_655}}
used the macroscopic Landau theory of phase transition, expanding
the free energy up to sixth power of the magnetization. If the
parameter $\mathcal{B}$ (with respect to M$^4$) is negative, the
transition can be first-order like. In this case, the
magnetization will present large field cycling irreversibility
\emph{only} for fields and temperatures below H$_c^*$ and T$_C^*$,
respectively. Further analysis of Landau theory, even for
$\mathcal{B}<$0 and higher temperatures and magnetic fields
(H>H$_c^*$ and T>T$_C^*$), show that the magnetization has a
peculiar inflexion point at a characteristic magnetic field H$_c$,
that increases linearly with temperature, H$_c$(T)$\propto$
(T-T$_0$). Mira \emph{et al.}
\cite{PRB_60_1999_2998,{PB_320_2002_23}} have also applied the
Landau theory to manganites. However, the results of the Landau
theory are not sufficient to reproduce all peculiar magnetic
properties of the manganites, such as the anomalous behavior of
the H/M vs. T quantity, presented on figure 5 and figure 12.

\subsection{Experimental and theoretical results for
La$_{0.60}$Y$_{0.07}$Ca$_{0.33}$MnO$_3$}\label{subsection_experimental_and_theoretical_data.}

A ferromagnetic ceramic La$_{0.60}$Y$_{0.07}$Ca$_{0.33}$MnO$_3$
was prepared by standard solid-state methods
\cite{JMMM_242_2002_655}, and the magnetization was measured using
a Quantum Design SQUID magnetometer (55 kOe) and an Oxford
Instruments VSM (120 kOe).

In this section we will apply the general results obtained in the
previous sections to a quantitative analysis of experimental data
for La$_{0.60}$Y$_{0.07}$Ca$_{0.33}$MnO$_3$. As stated in section
\ref{section_Mean-Field_Approximation}, we work within the
mean-field approximation, for which $x$ assume the expression:
\begin{equation}\label{x_mean_field}
  x=\frac{\mu (H+\lambda \mathcal{M}_q)}{kT}
\end{equation}
where $\lambda$ is the mean-field parameter. Figure 10 presents
the measured and theoretical magnetic moment as a function of
magnetic field, for several temperatures above T$_C$. The
excellent experimental-theoretical agreement is due to the use of
the Tsallis statistics, which parametrizes the system
inhomogeneity \cite{PR_344_2001_1}.

Here, we use $q$, $\mu$, $\lambda$ and $N$, the number of clusters
in the sample, as free parameters. The magnetic moment $\mu$ of
the clusters follows the usual temperature tendency of a
para-ferromagnetic transition, whereas $q$ increases towards unity
with increasing temperature (figure 11). The temperature
dependence of the $q$ parameter was expected, since it should
reach the unity for sufficiently high temperatures.

From these results, we were able to compare the anomalous downturn
in H/M vs. T curves, just with the temperature dependence of $q$
and $\mu$, and the approximately constant values of $\lambda$ and
$N$. The calculated H/$\mathcal{M}_q$ vs. T curve is displayed in
figure 12, with its corresponding experimental value. One should
stress that the solid line in this plot does not include any other
fitting parameter. The curve was calculated using only the fitted
parameters obtained from Figure 10.

Additionally, Amaral and co-workers pointed out that
La$_{0.60}$Y$_{0.07}$Ca$_{0.33}$MnO$_3$ \cite{JAP_sem_numero}
presents two different features on the inflection point, at $H_c$,
of its $M$ vs. $H$ curves: (i) a linear temperature dependence of
the characteristic field $H_c$, for T$>$T$_C^*$, and (ii) an
observed hysteresis for temperatures in the range
T$_C<$T$<$T$_C^*$. Figure 13 shows the plot of $H_c$ vs. $T$
obtained from experimental data. It is striking the similarity
between this plot and the curve shown in Fig. 7(c).

Finally, the Arrot Plots presented by Amaral \emph{et al.}
\cite{JAP_sem_numero,{JMMM_242_2002_655}} and Mira \emph{et al.}
\cite{PRB_60_1999_2998,{PB_320_2002_23}} are very similar to those
presented on figure 9(a)(b).

\section{Conclusion}

In two previous publications
\cite{EL_58_2002_42,{PRB_66_2002_134417}} we presented evidences
that the magnetic properties of manganites can be suitably
described in the framework of Tsallis statistics. In the present
paper we have extended our analysis and presented new compelling
evidences in this direction by deducing a magnetic phase diagram
which matches observed experimental results on
La$_{0.60}$Y$_{0.07}$Ca$_{0.33}$MnO$_3$, along with some other
magnetic properties of this compound. The interpretation of the
entropic parameter $q$, given by Beck and Cohen
\cite{cond-mat/0205097}, in terms of the ratio between the mean
and width of the temperature distribution in the system, comes to
support our proposal, since manganites have long been recognized
as objects whose properties are dominated by intrinsic
inhomogeneities \cite{PR_344_2001_1,{JMMM_200_1999_1},{NMR_prca}}.
Such a distribution can be translated as a distribution of the
magnetic susceptibility, as pointed out by Salamon \emph{et al.}
\cite{PRL_88_2002_197203}, from which a temperature dependence of
$q$ can be expected. Therefore, we conclude that the Tsallis
nonextensive statistics is a handy tool to study, classify and
predict magnetic and thermal properties of manganites.

\section{Acknowledgements}

The authors thanks FAPERJ/Brasil, FCT/Portugal (contract
POCTI/CTM/35462/99) and ICCTI/CAPES (Portugal-Brasil bilateral
cooperation), for financial support. We are also thankful to A.P.
Guimarães, E.K. Lenzi and R.S. Mendes, for their helpful
suggestions.

\section{Figure Captions}
\begin{small}
\textbf{Figure 1.} Gibbs free energy(Eq.\ref{gibbs}), within the
mean field approximation and $h=$0, as a function of the reduced
magnetization. (a) For $q>$0.5 only one minimum is observed, for
any value of temperature (t$>$t$_c$ or t$<$t$_c$), representing a
second order phase transition for the magnetization. (b) For
$q<$0.5 and t$=$t$_c$ two degenerated minima are observed,
representing a first order phase transition. The t$_{SH}$ and
t$_{SC}$ temperatures limit the 'superheating' and 'supercooling'
cycle, responsible for the thermal hysteresis normally observed in
first order transitions.

\textbf{Figure 2.} Temperature dependence of the reduced
equilibrium magnetization for several values of $q$ and $h=0$. For
$q<$1/2 the transition shows a first order character, whereas for
$q>$1/2 the transition is of second order type. Inset: Thermal
hysteresis normally observed in first order phase transitions,
with the 'superheating' and 'supercooling' cycles.

\textbf{Figure 3.} Temperature dependence of the generalized
magnetic susceptibility $\chi _q$, and its inverse, for (a)
$q<$0.5 and (b) $q>$0.5.

\textbf{Figure 4.} Temperature dependence of the reduced
equilibrium magnetization for several values of $h$, and $q=$0.1.
Sufficiently high magnetic field $h\geq h_{0q}$ (Eq.\ref{b0}), is
able to remove the discontinuity in the equilibrium magnetization
curve.

\textbf{Figure 5.} Temperature dependence of the quantity
h/$\mathcal{M}_q$, for $q=$0.2 and $h=3h_{0q}$.

\textbf{Figure 6.} Magnetic field dependence of the reduced
equilibrium magnetization for several values of temperature, and
$q=$0.1. It represents a genuine ferromagnetic phase transition
induced by magnetic field, in the paramagnetic phase. For
sufficiently high temperatures $t\geq t_{0q}$ (Eq.\ref{t0}), the
transition becomes continuous.

\textbf{Figure 7.} $q-$magnetic phase diagram summarizing the main
properties found in the model. (a) Projection of the phase diagram
in $t-q$ plane. The solid lines divide the plane in ferro (below)
and paramagnetic (above) phases, with a region of phase
coexistence between the dotted lines (shaded area). On the left
side of the dashed line, the transition has a first order
character, whereas a second order character arises for the right
side. (b) Projection of the phase diagram in $h-q$ plane, for
several temperatures t$>$t$_c$. The curves in this case have a
complete analogy to the previous one, however, it is ferromagnetic
above the transition lines, since it represents a ferromagnetic
transition induced by magnetic field. (c) Projection of the phase
diagram in the $h-t$ plane, for $q=$0.1. Above certain values of
field and temperatures $h_{0q}$ (Eq.\ref{b0}), and $t_{0q}$
(Eq.\ref{t0}), the transition becomes second order.

\textbf{Figure 8.} The generalized Landau coefficient
$\mathcal{B}_q$ as a function of the entropic parameter $q$. For
$q<$0.5, $\mathcal{B}_q$ assume negative values, indicating first
order phase transition. Inset: temperature dependence of the
parameter $\mathcal{A}_q$, that represents the inverse of the
generalized susceptibility.

\textbf{Figure 9.} The generalized Arrot Plot (h/m vs. m$^2$
curves), for (a) $q>$0.5 and (b) $q<$0.5.

\textbf{Figure 10.} Measured (open circles) and theoretical (solid
lines - Eqs.\ref{langevin_general} and \ref{x_mean_field})
magnetic moment as a function of magnetic field, for several
values of temperatures above T$_C$=150 K.

\textbf{Figure 11.} Temperature dependence of the fitting
parameters, $q$ and $\mu$ (see text).

\textbf{Figure 12.} Measured (open circles) and theoretical (solid
lines) values of the quantity H/M vs. T. The solid line in this
plot does not include any fitting parameters, and was calculated
using only the fitted parameters obtained from figure 10.

\textbf{Figure 13.} The linear temperature dependence, for
T$>$T$_C^*$, of the characteristic field H$_c$, which corresponds
to the inflexion point of the experimental M vs. H curves,
measured in La$_{0.60}$Y$_{0.07}$Ca$_{0.33}$MnO$_3$. For
T$_C<$T$<$T$_C^*$ the hysteresis is indicated by the shaded area.
It is striking the similarity between this experimental plot and
the theoretical one, shown in figure 7(c).

\end{small}

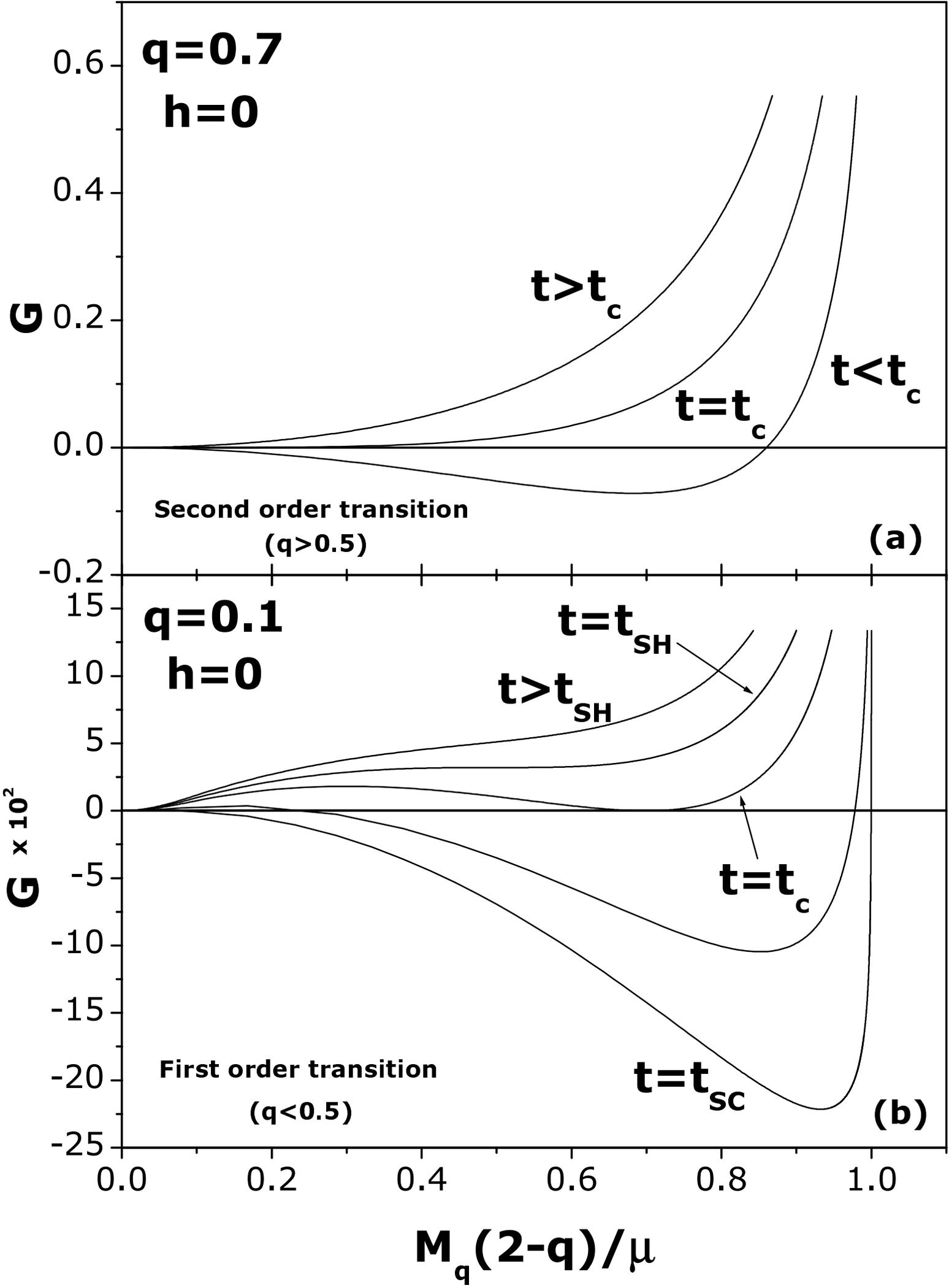
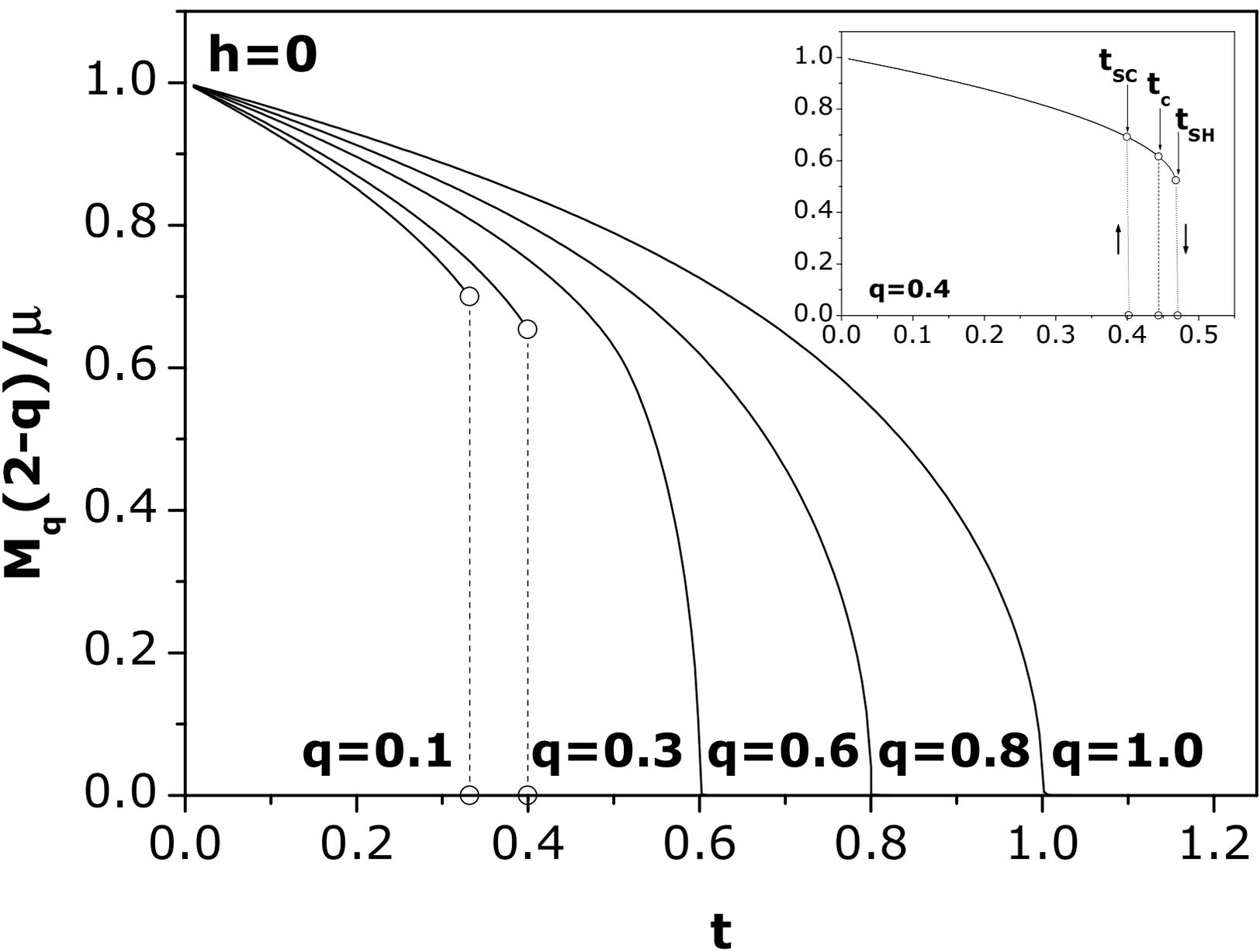
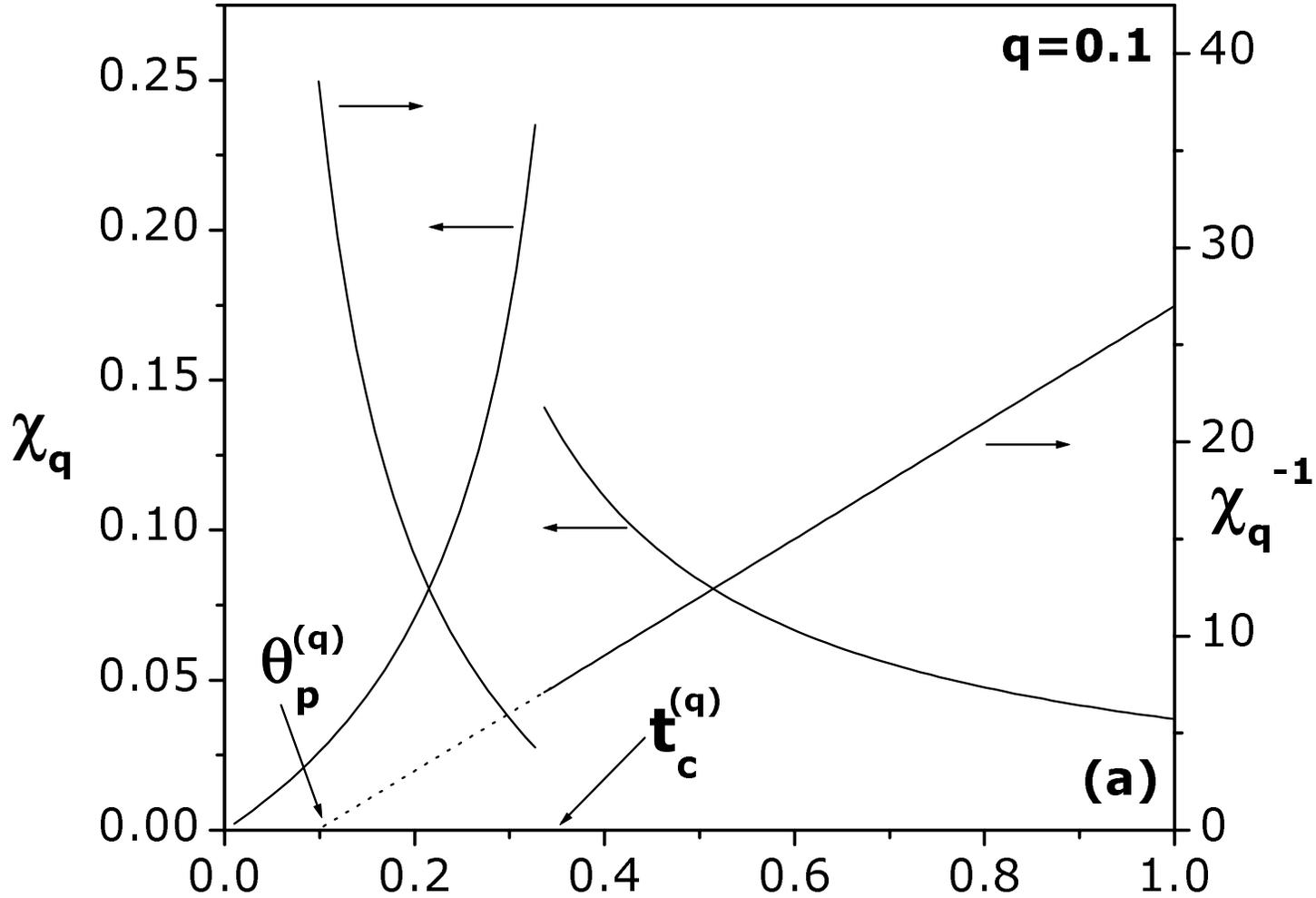
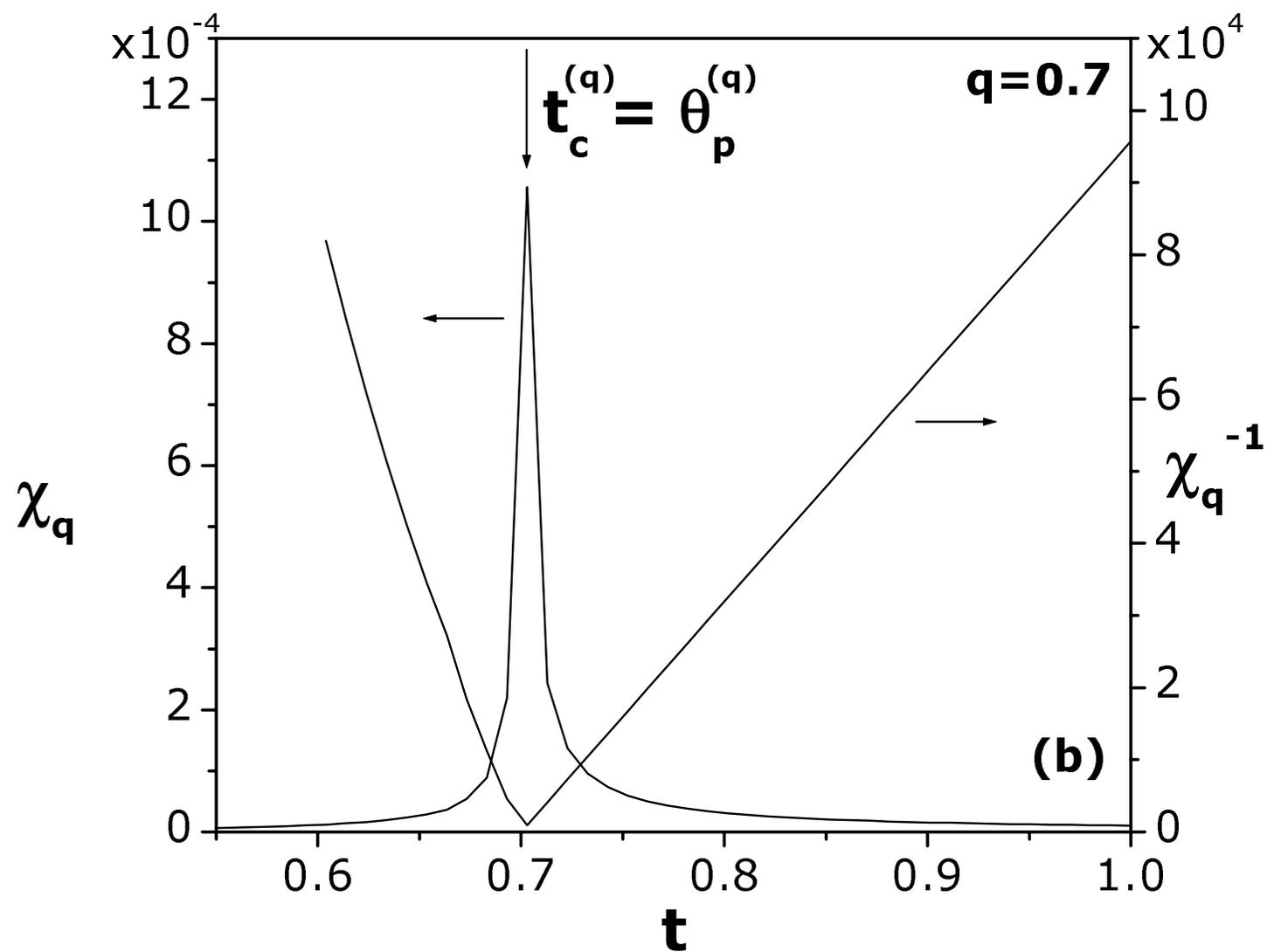
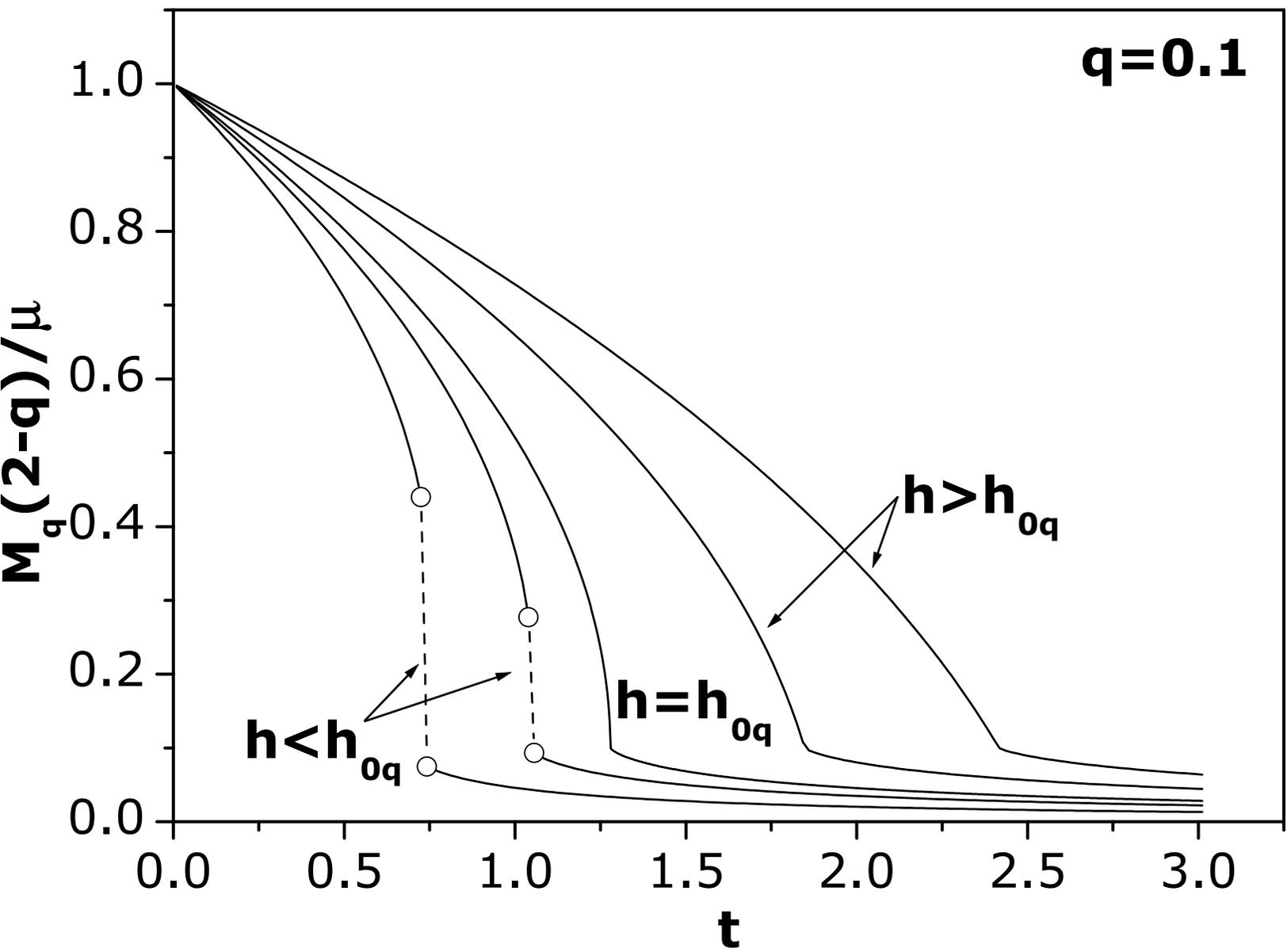
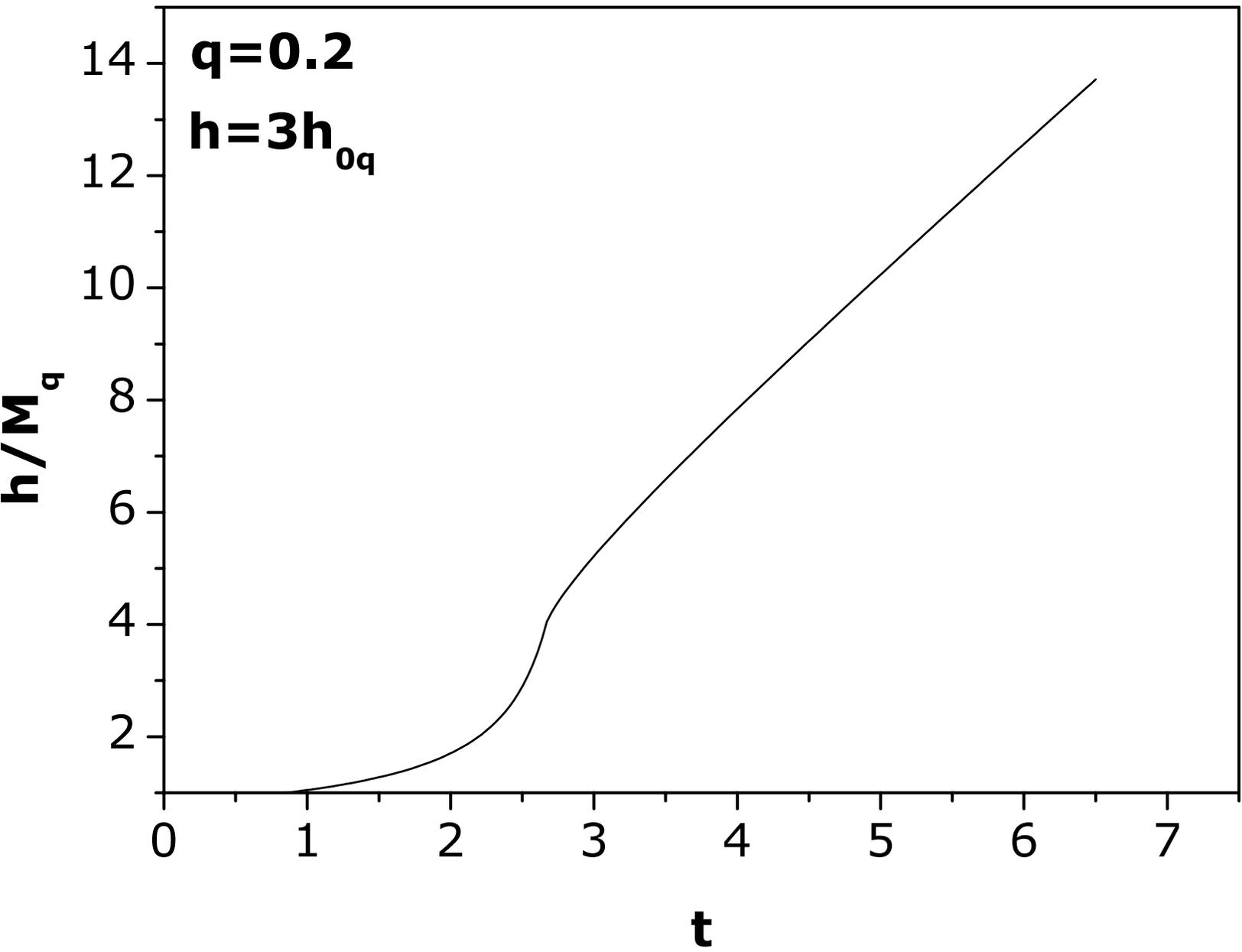
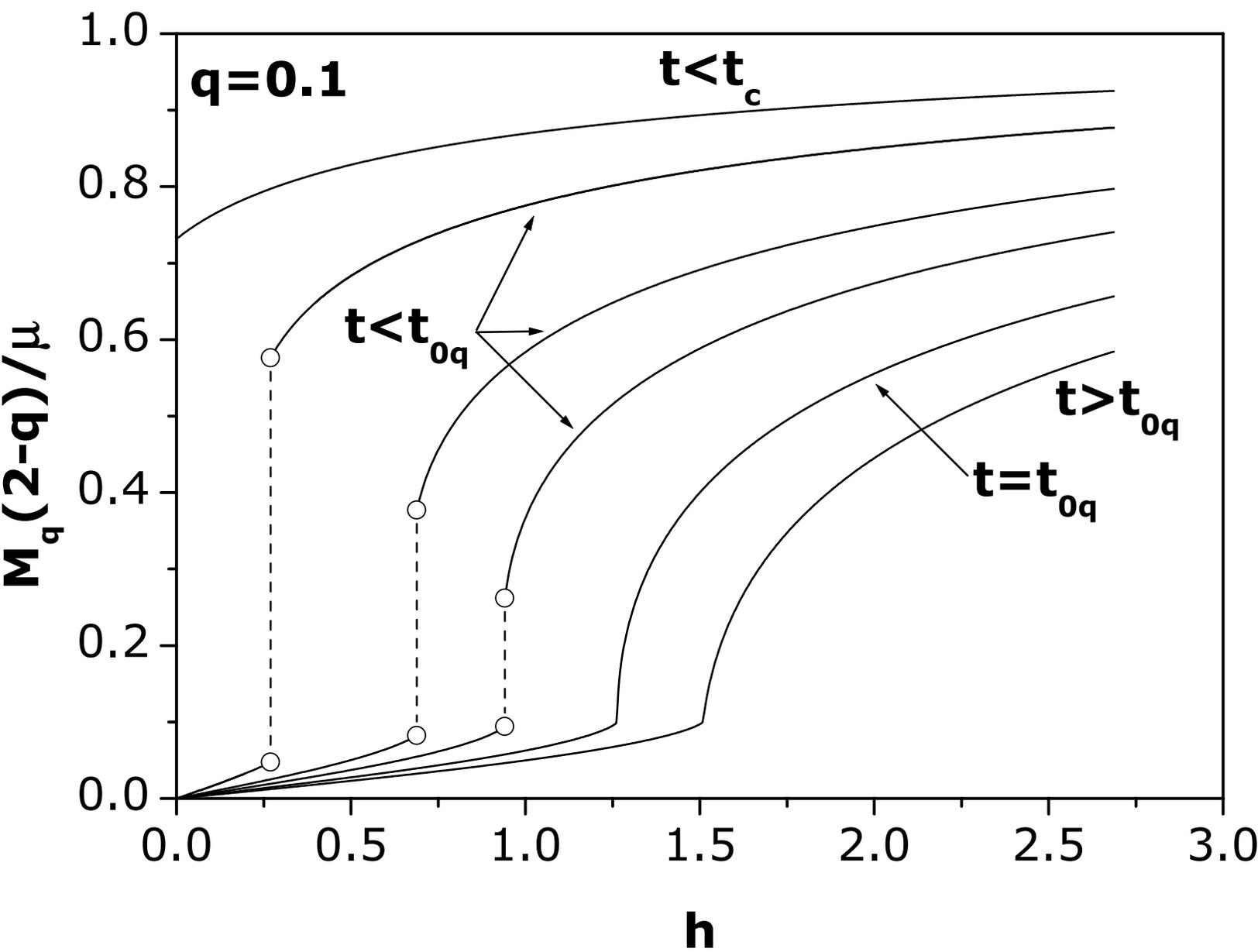
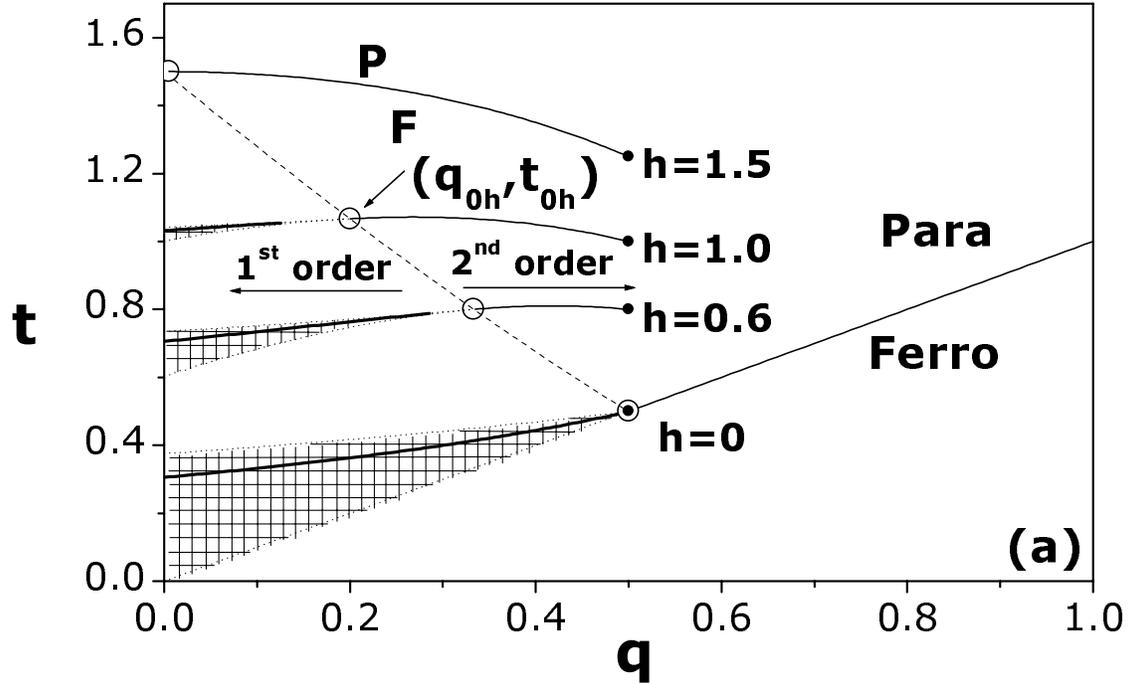
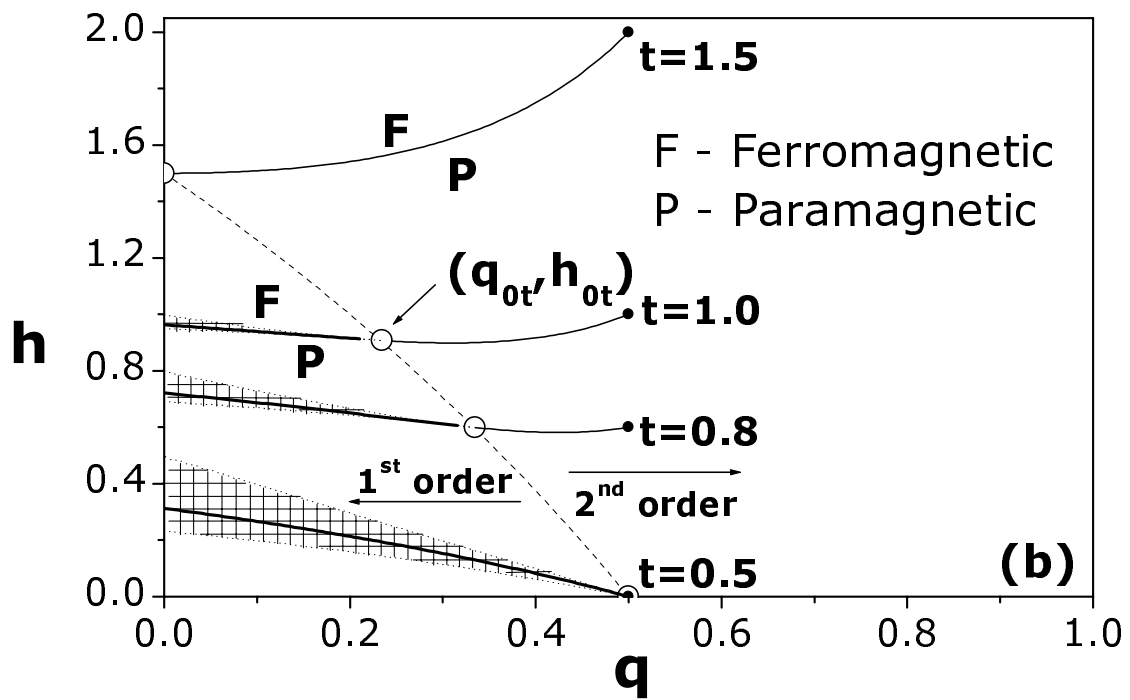
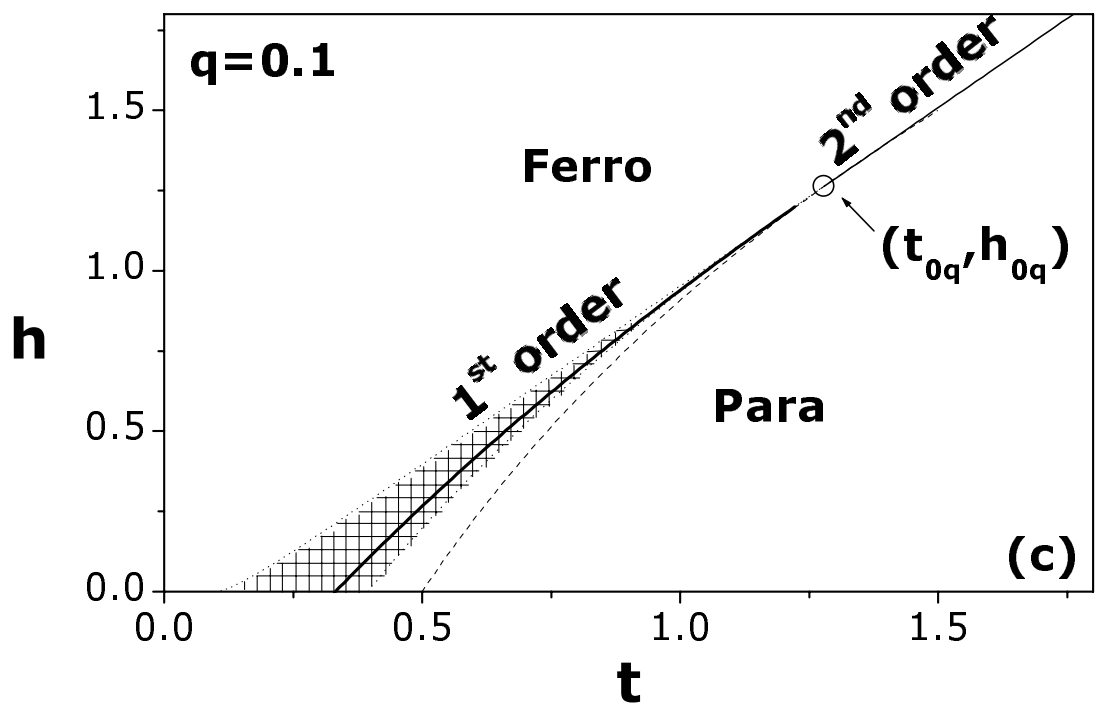
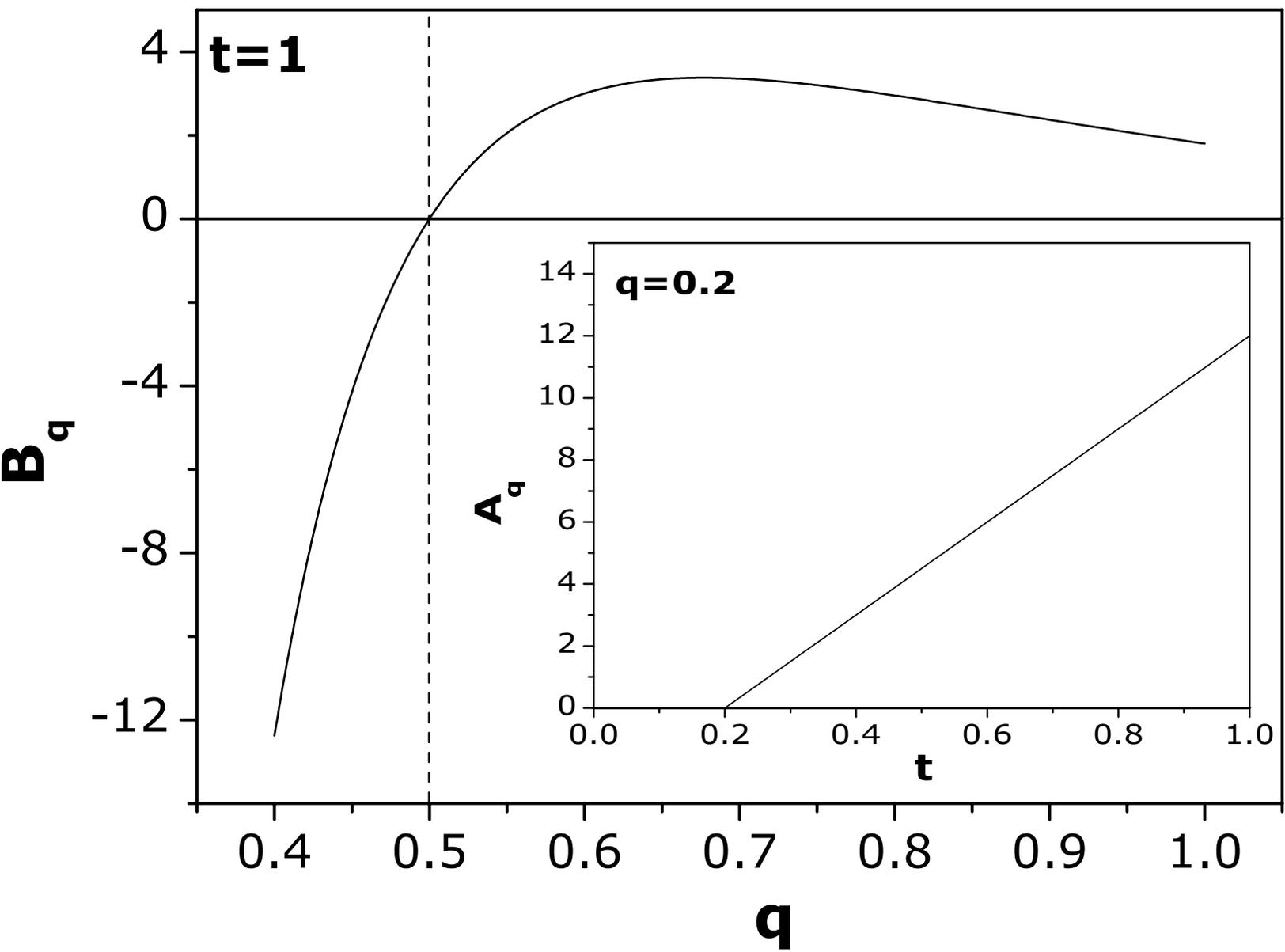
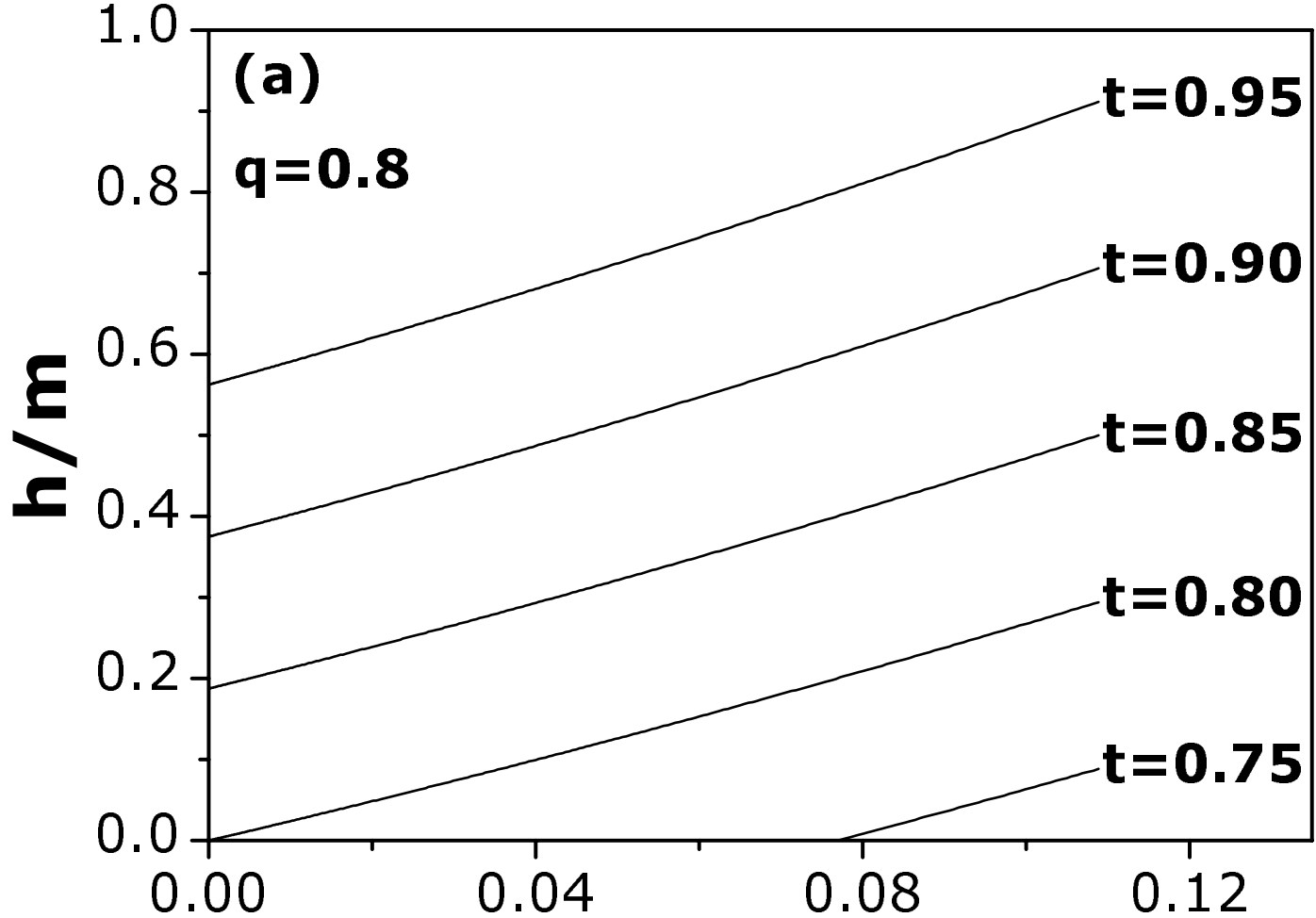
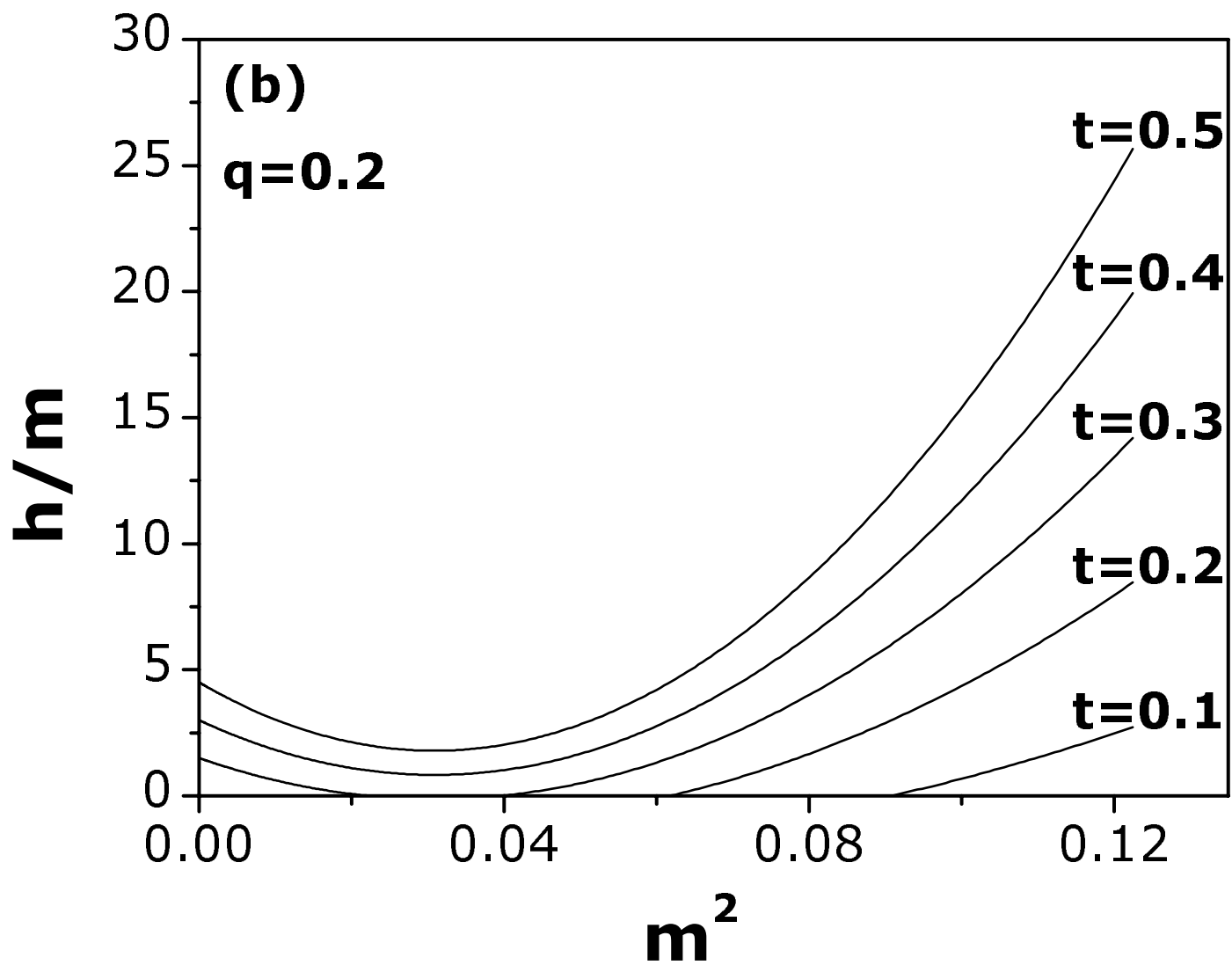
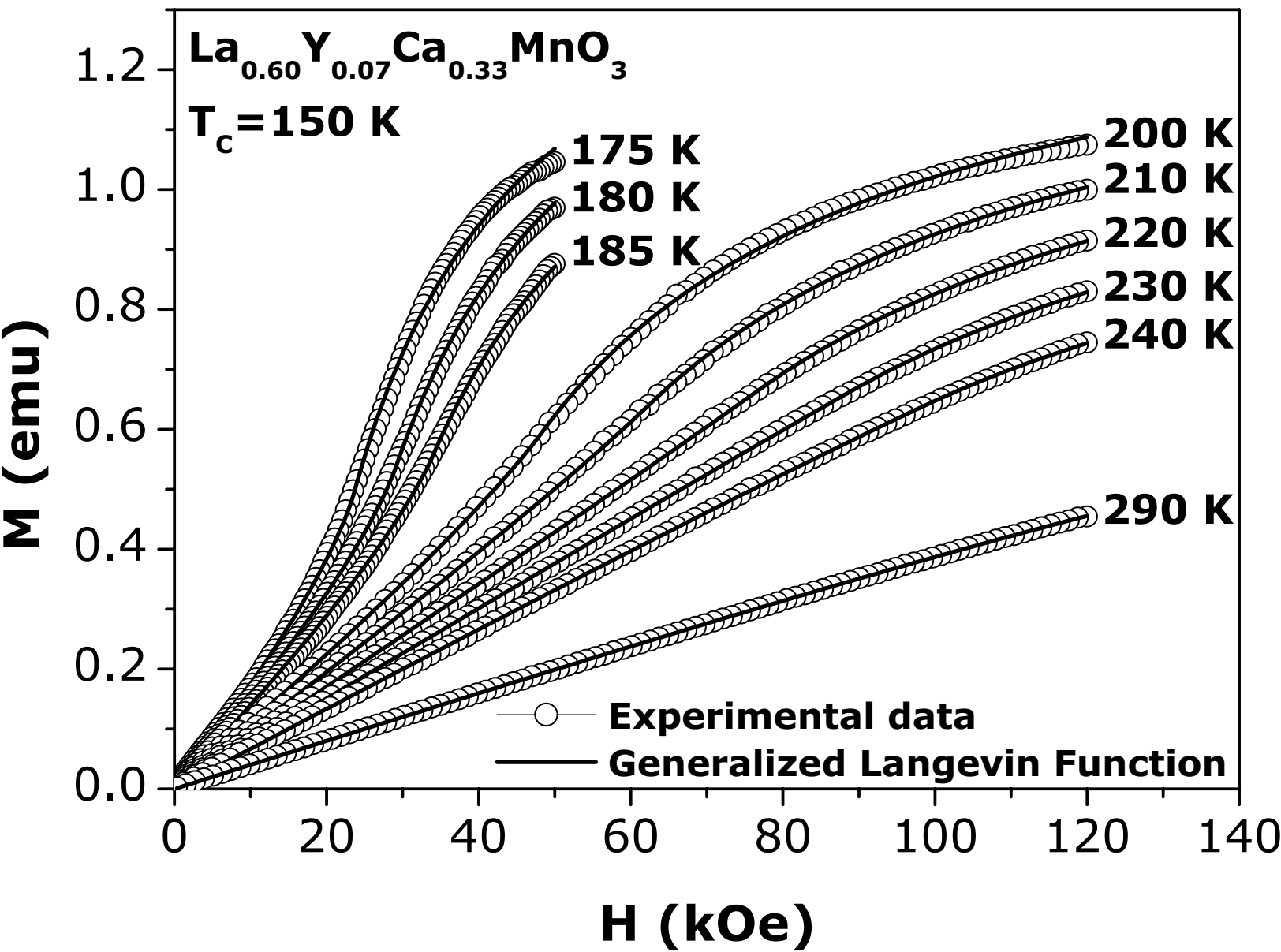
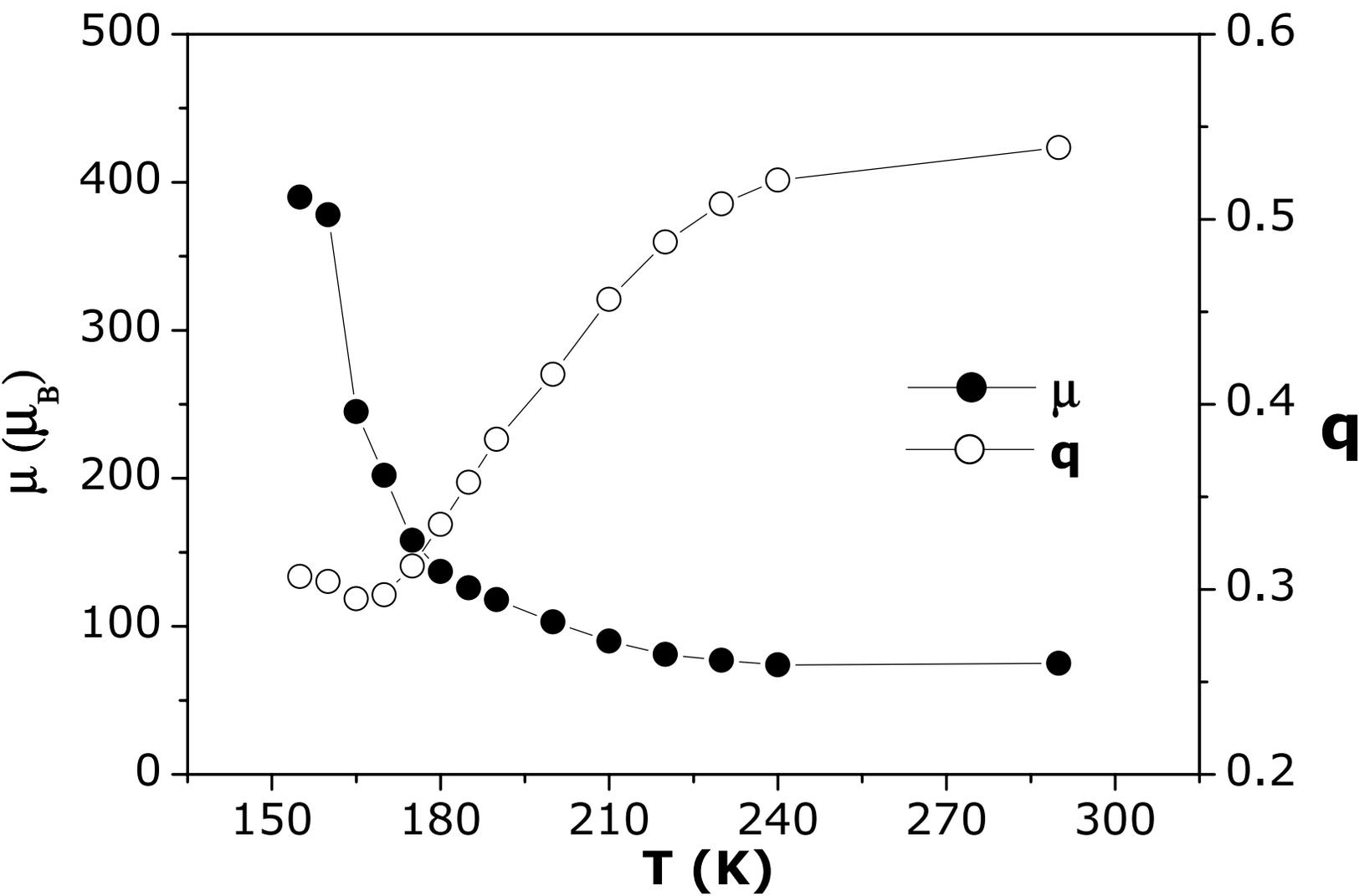
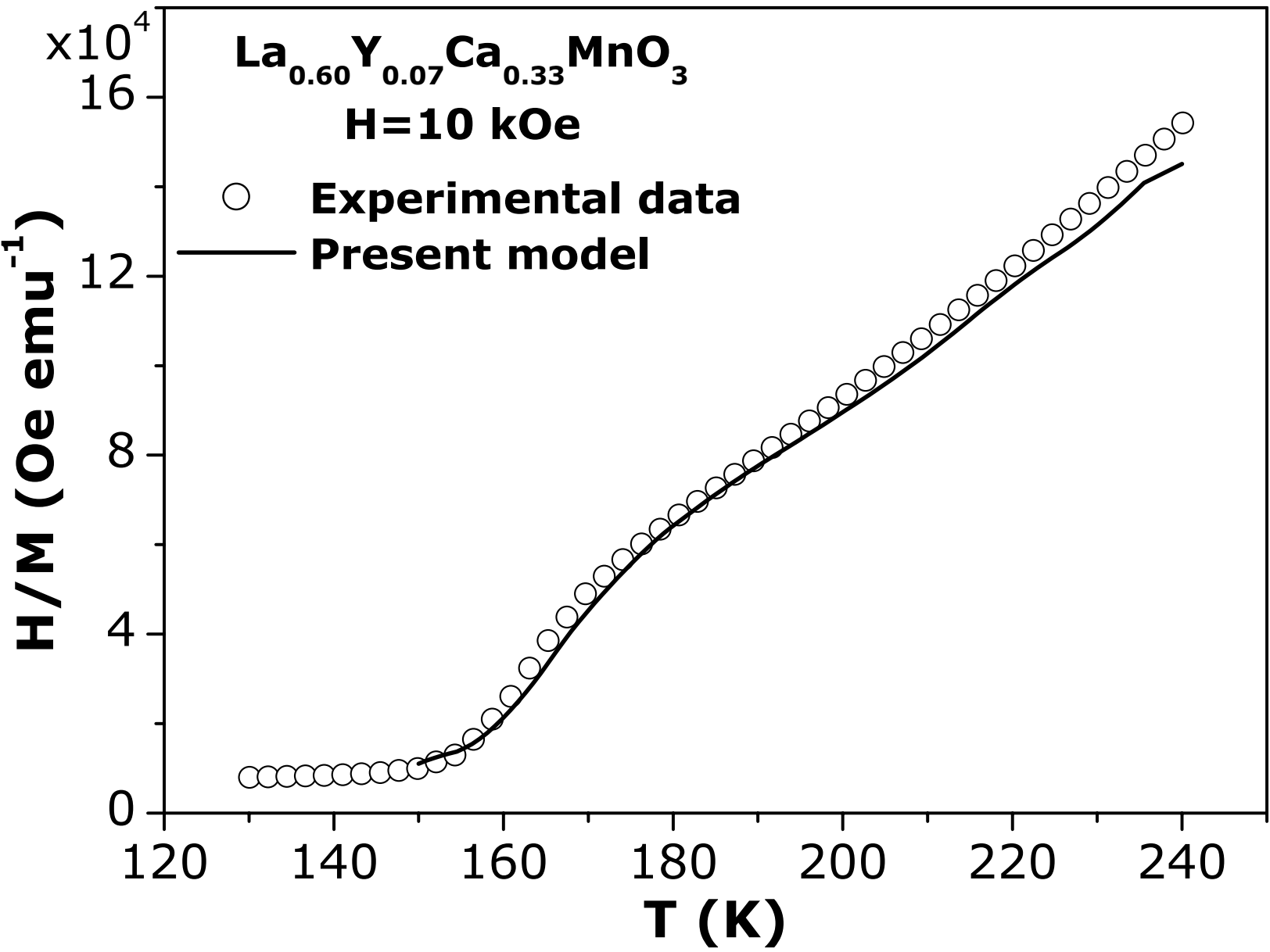
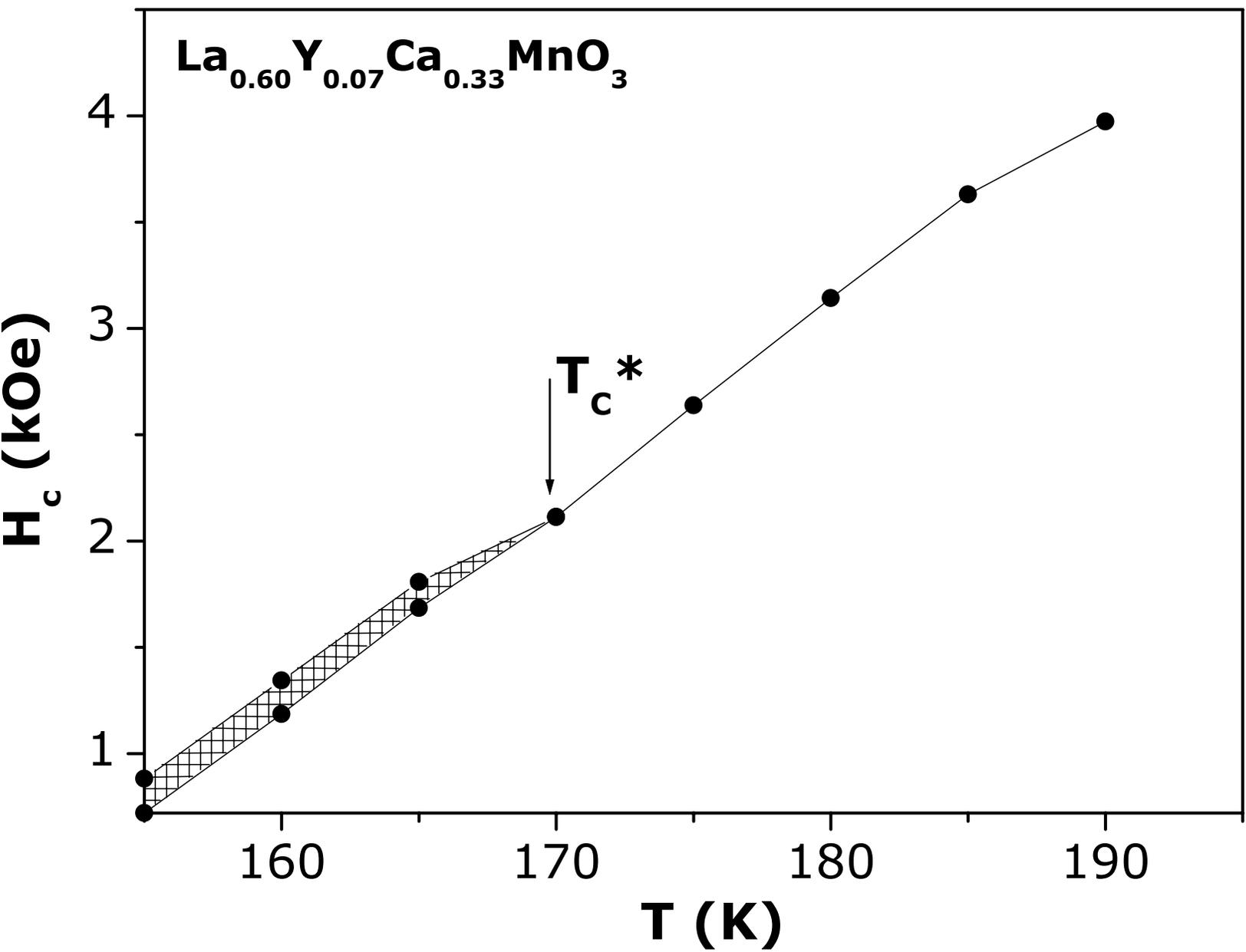
\end{document}